# The broadband pulse diffraction on a half-plane screen and caustic

P.A. Golovinski [1,2), V.A. Astapenko [1)

[1) *Moscow Institute of Physics and Technology (State University), 141700, Moscow, Russia*

[2) *Physics Research Laboratory, Voronezh State University of Architecture and Civil Engineering, 394006, Voronezh, Russia*

*e-mail: golovinski@bk.ru*

The diffraction of ultrashort pulse changes its spatial and temporal structure that is crucial for multi-channel communication and location via such pulses. The features of the evolution of broadband pulses discussed for two general problems: diffraction at the edge of a flat absorbing screen and the field near a simple caustic. The basic equations are formulated for spatial distribution of the pulse field after the screen in the spectral and time domain. Calculations of pulse diffraction for different wave shapes and comparison of results are presented. We obtain an integral representation for pulse field with arbitrary initial form near a caustic and analyze the results for different waveforms.

## 1. Introduction

Classical diffraction theory [1, 2] formulates generic foundations for solution both steady state and time dependent problems via the Kirchhoff-Somerfield integral equation. Contemporary approach to the diffraction is mainly applied as the physical diffraction theory of monochromatic waves [3]. However, the development technology of broadband radio frequency waves and ultrashort optical pulses has formulated the problem to expend theory to such malty-frequency signals.

When diffraction of short pulses a new phenomenon is observed that is absent in the case of monochromatic or quasi-monochromatic waves [4, 5]. In particular, the narrow aperture induces differentiation effect of the initial pulse, increasing the number of nodes in it to unity. This phenomenon has radically changed the structure of space-time pulse after passing the narrow holes [6]. A similar effect occurs during the passage ultrashort pulses of differentiation diffraction gratings, which allows obtain the desired shape of the low-cycle pulses [7]. The effect of differentiating single-cycle terahertz pulse by passing the focus has been experimentally observed and confirmed by numerical calculations [8, 9].

As broadband pulses are increasingly used in a variety of applications, including communication and radar, the study features of the diffraction associated with a wide spectrum of the signal is of considerable interest. In this paper we consider the problem of diffraction of a wideband signal on the straight edge of the semi-infinite screen and the structure of the field in the caustic.

## 2. Diffraction at the edge of the screen

Let the incident pulse propagates from a remote source, so it can be considered as a point source, and the spherical wave about plane. We take a convenient coordinate system, in which the XY plane passes through the source and the edge of the screen. It is perpendicular thereto XZ plane so that it passes through the point source position S and the observation point P perpendicular to the plane XY. The intersection of these planes with the line edge of the screen gives the point of origin O [10], as shown in Fig.1.

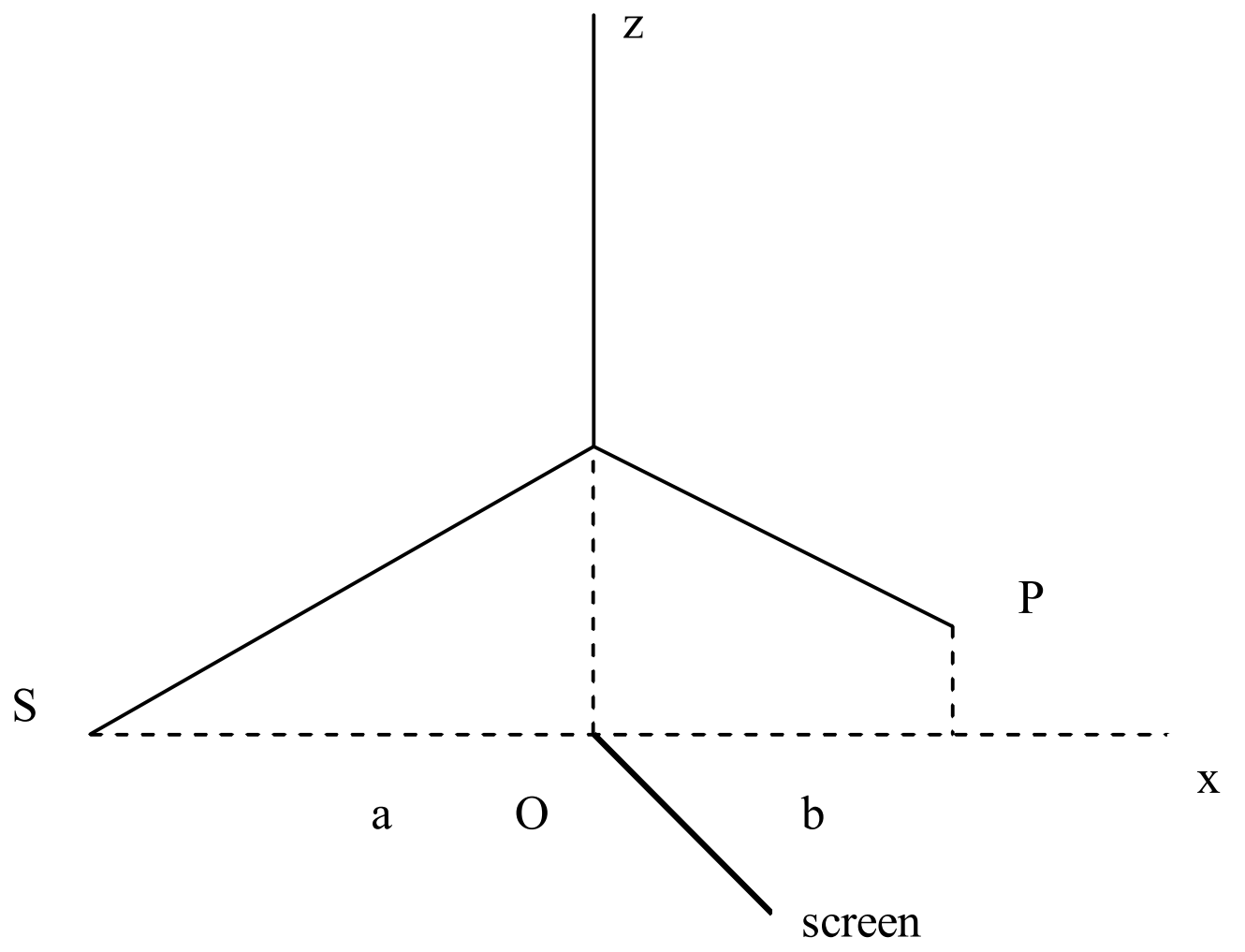

**Fig.1.** The relative position of diffracted ray and the edge of the screen.

The field at the observation point, that is produced by monochromatic wave, in the scalar approximation and in the model of a completely absorbing screen, is obtained according the Huygens-Fresnel principle [11]. It is expressed in terms of Fresnel integrals [12] in the form

$$u_P(\mathbf{r},\omega) = \frac{A(1+i)}{2(a+b)} \exp[ik(a+b) + isb/a] \times \left[\left(\mathrm{C}(s) + \frac{1}{2}\right) + i\left(\mathrm{S}(s) + \frac{1}{2}\right)\right]. \tag{1}$$

Here $a$ is a distance from the source S to the point O, $b$ is $x$-coordinate of the observation point P. The other parameters are given by

$$s = z_P \sqrt{\frac{ka}{2b(a+b)}}, k = \frac{\omega}{c}, \tag{2}$$

$$\mathrm{C}(s) = \sqrt{\frac{\pi}{2}} \int_0^s \cos t^2 \, dt, \ \mathrm{S}(s) = \sqrt{\frac{\pi}{2}} \int_0^s \sin t^2 \, dt.$$

For the free propagation of the wave field at P would be

$$u_0(\mathbf{r},t) = \frac{A}{a+b} e^{ik(a+b)}. \tag{3}$$

It is clear that variations in the distribution of the field behind the screen are determined by the respective Fresnel integrals. The spectral distribution of the initial pulse is described by function

$$\tilde{u}_0(0,\omega) = \frac{1}{2\pi} \int_{-\infty}^{\infty} u_0(0,t) e^{-i\omega t} \, dt, \tag{4}$$

and the square of the spectral components of the diffracted pulse is characterized by its spectral energy density [13]

$$|\tilde{u}_0(\mathbf{r},\omega)|^2 = \frac{1}{2} \left| \frac{\tilde{u}_0(\omega)}{(a+b)} \right|^2 \left| \left(\mathrm{C}(s) + \frac{1}{2}\right) + i\left(\mathrm{S}(s) + \frac{1}{2}\right) \right|^2. \tag{5}$$

With decreasing frequency, the width of the penumbra behind the edge of the screen increases, and when, $\omega \to 0$, $z_P \to \infty$. This means that when calculating the physical effects of diffraction of broadband signals, it is important to have a correct description of the spectrum. It manifested only frequencies in a finite range around the carrier frequency capturing frequency is not close to zero. Accordingly, the use of Gaussian pulses in the description of the phenomena of diffraction broadband pulses should be with caution, to avoid making artifacts.

To illustrate the role of the spectral width in Fig. 3 shows the results of calculation for the diffraction of a pulse with a Gaussian spectrum and pulse with the rectangular shape of the spectral function as presented in Fig. 2.

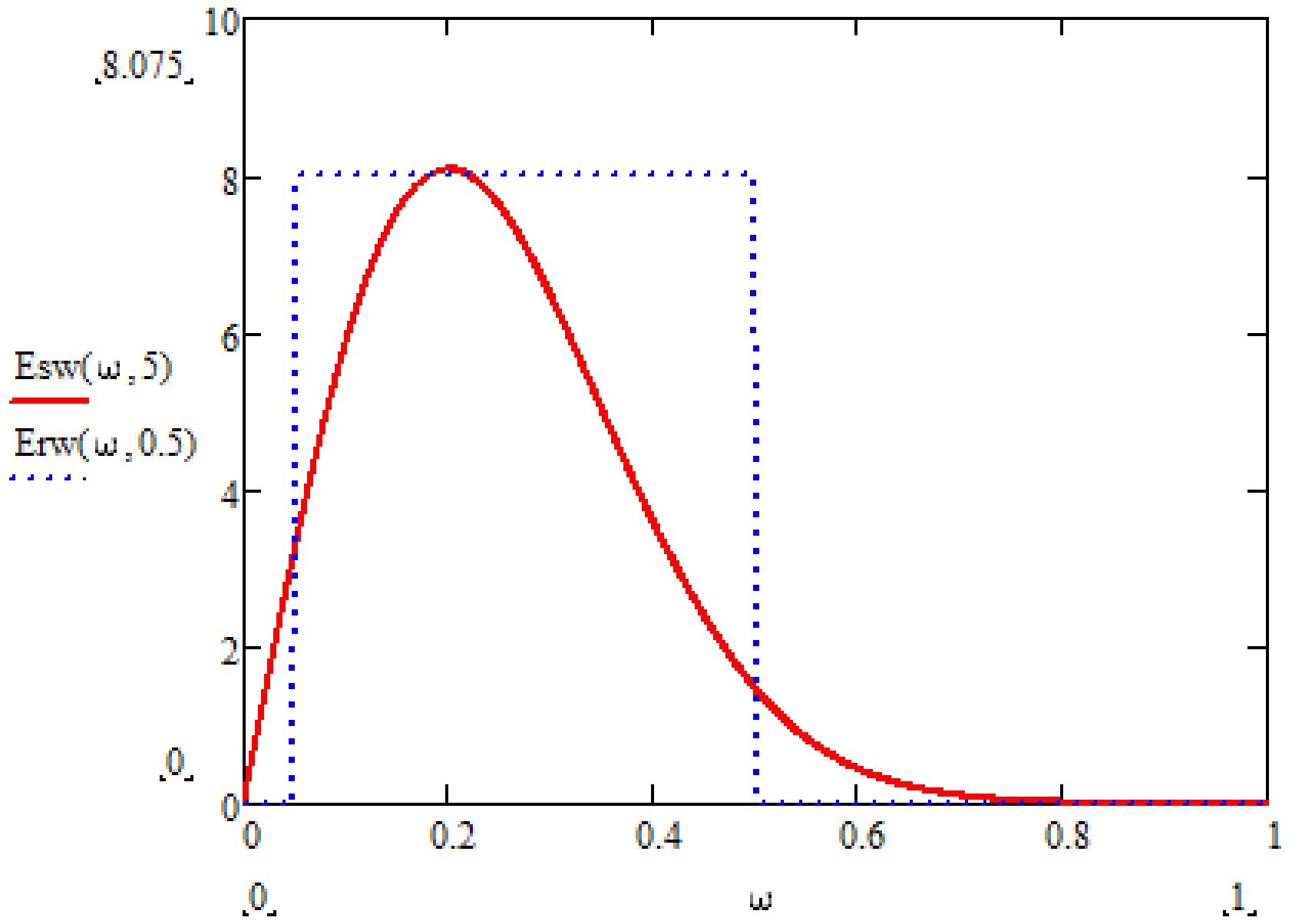

**Fig.2.** Gaussian pulses and rectangular shape of the spectral function.

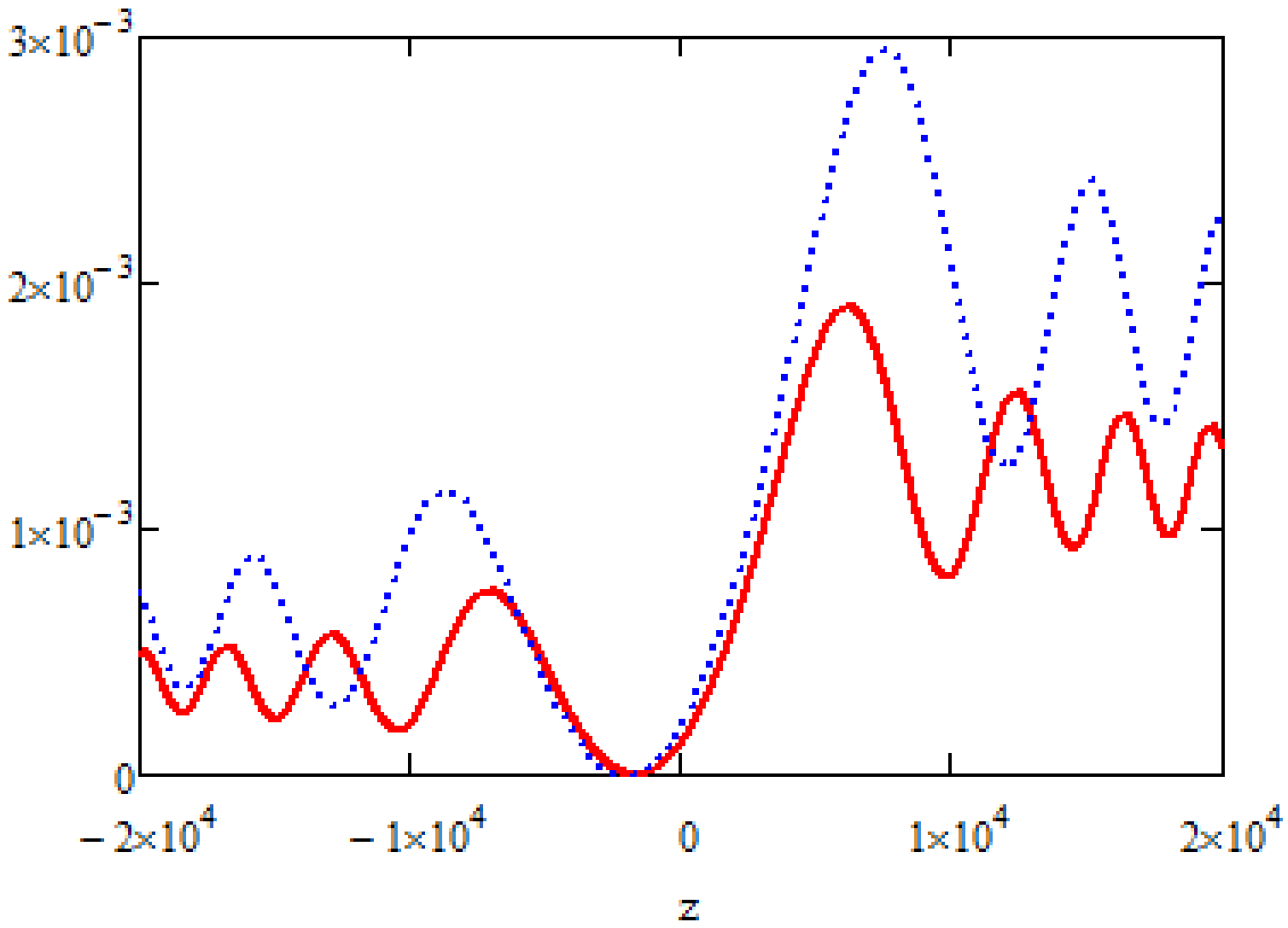

**Fig.3.** The dependence of the spectral energy density on the pulse frequency and the coordinates of the observation point displacement from the edge of the screen: solid line – ω=0.03 at.u., dotted line – ω=0.02 at.u. Ordinate is in arb. u., abscissa is in at. u., $a=b=10^3$ .

### 3. Temporary diffraction pattern of the pulse on the edge

Let au consider the problem of space-time structure of the pulse is diffracted in the region. The solution can be obtained by performing the inverse Fourier transform for the spectral components of the diffracted pulse. We can also directly solve the dynamic problem [14]. In the second case it is convenient to use unsteady Kirchhoff-Sommerfeld integral in the form

$$u_P(\mathbf{r},t) = \frac{1}{4\pi}\iint_S \left[ u\frac{\partial}{\partial n}\left(\frac{1}{r}\right) - \frac{1}{cr}\frac{\partial r}{\partial n}\left(\frac{\partial u}{\partial t}\right) - \frac{1}{r}\left(\frac{\partial u}{\partial n}\right)\right] dS \tag{6}$$

as the response to the wave in $\delta$-pulse form [5, 15-17]. The corresponding to the pulse $\delta(t - x/c)$ response function has the form

$$g(\mathbf{r},t) = \frac{z_P}{r^3}\delta(t - x/c) + \frac{1}{cr}\left[1 + \frac{z_P}{r}\right]\delta'(t - x/c), \tag{7}$$

where $r$ is the distance from the elementary portion of the surface of integration to the observation point, $c$ is the speed of light.

For the problem of diffraction $\delta$-pulse for normal incidence on the half-plane integration over the half-plane leads to a response function [16]

$$G(\mathbf{r},t) = \begin{cases} c\delta(t - x/c) + \dfrac{c}{2\pi}\dfrac{z_P\theta(t - x/c)}{(x - ct)\sqrt{(ct)^2 - r^2}}, z_P > 0, \\ \dfrac{c}{2\pi}\dfrac{z_P\theta(t - x/c)}{(x - ct)\sqrt{(ct)^2 - r^2}}, z_P < 0. \end{cases} \tag{8}$$

The exceptional simplicity of this formula makes it very convenient for the calculation of the diffraction of short pulses. Diffraction pulse of arbitrary shape can be found by calculating the integral

$$u_P(\mathbf{r},t) = \int_{-\infty}^{\infty} G(\mathbf{r}, t - \tau) u_0(0,\tau)\, d\tau. \tag{9}$$

### 4. Field near the caustic

In the center of curvature of the wave surfaces in the geometric optics approximation the radiation intensity accesses formally to infinity. It is therefore of special interest the field behavior at the centers of curvature of the wave surfaces calculated from the wave theory.

The distribution of the harmonic field amplitude near the caustic is described by the integral formula [10]

$$u_P(x,\omega) = A\int_{-\infty}^{\infty} \exp\left(-ikx\theta - i\frac{k\rho}{6}\theta^3\right) d\theta. \tag{10}$$

It is actually the integral representation of the Airy function [12], so that $u_p \sim \mathrm{Ai}\left(x\sqrt[3]{2k^2/\rho}\right)$. Here $\rho$ is radius of curvature of the caustic, $x$ is distance from the contact point of the beam with the caustic to the observation point P with respect to the normal, $\theta$ is angle of rotation of the caustic radius at which the integration is performed.

For a pulse with a known form $u_0(0,t)$ we have

$$u_P(x,t) = \int_{-\infty}^{\infty} u_0\left(0, t + \frac{x}{c}\theta + \frac{\rho}{c6}\theta^3\right) d\theta . \qquad (11)$$

Equation (11) is a convenient integral representation of the pulse field in the vicinity of the caustic.

When generating ultra-wideband pulses are obtained single-cycle pulses and pulses of a few oscillations in radio frequency range [18, 19] and by diode generation [20]. Analogous pulse shape can be obtained in the optical [21] and THz [22] range. In view of the identified characteristic of temporary forms of ultrashort pulses of radio and optical range, convenient form ultrashort pulse description is wavelets in the form of Gaussian function $E_0(\tau) = A_1 \exp\left(-\tau^2\right)$ and its derivatives:

$$E_1(\tau) = A_2\left\{-\tau \exp\left(-\tau^2\right)\right\}, \qquad (12)$$
$$E_2(\tau) = A_2\left\{\left(1 - 2\tau^2\right)\exp\left(-\tau^2\right)\right\}.$$

Here $\tau = t/T$, $T$ is a quantity characterizing the pulse duration. The form of these pulses is shown in Fig. 4 a) and b).

a)

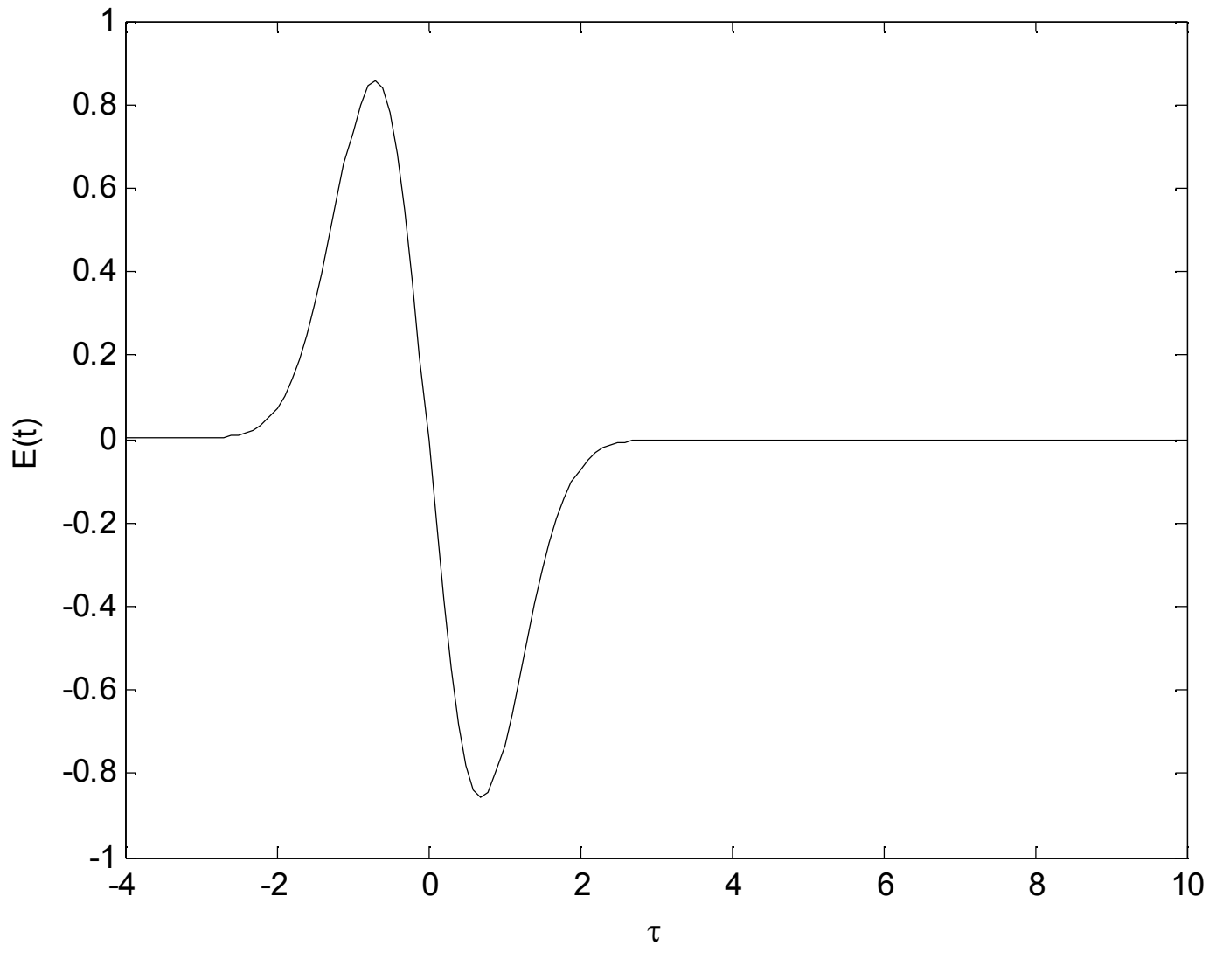

**в)**

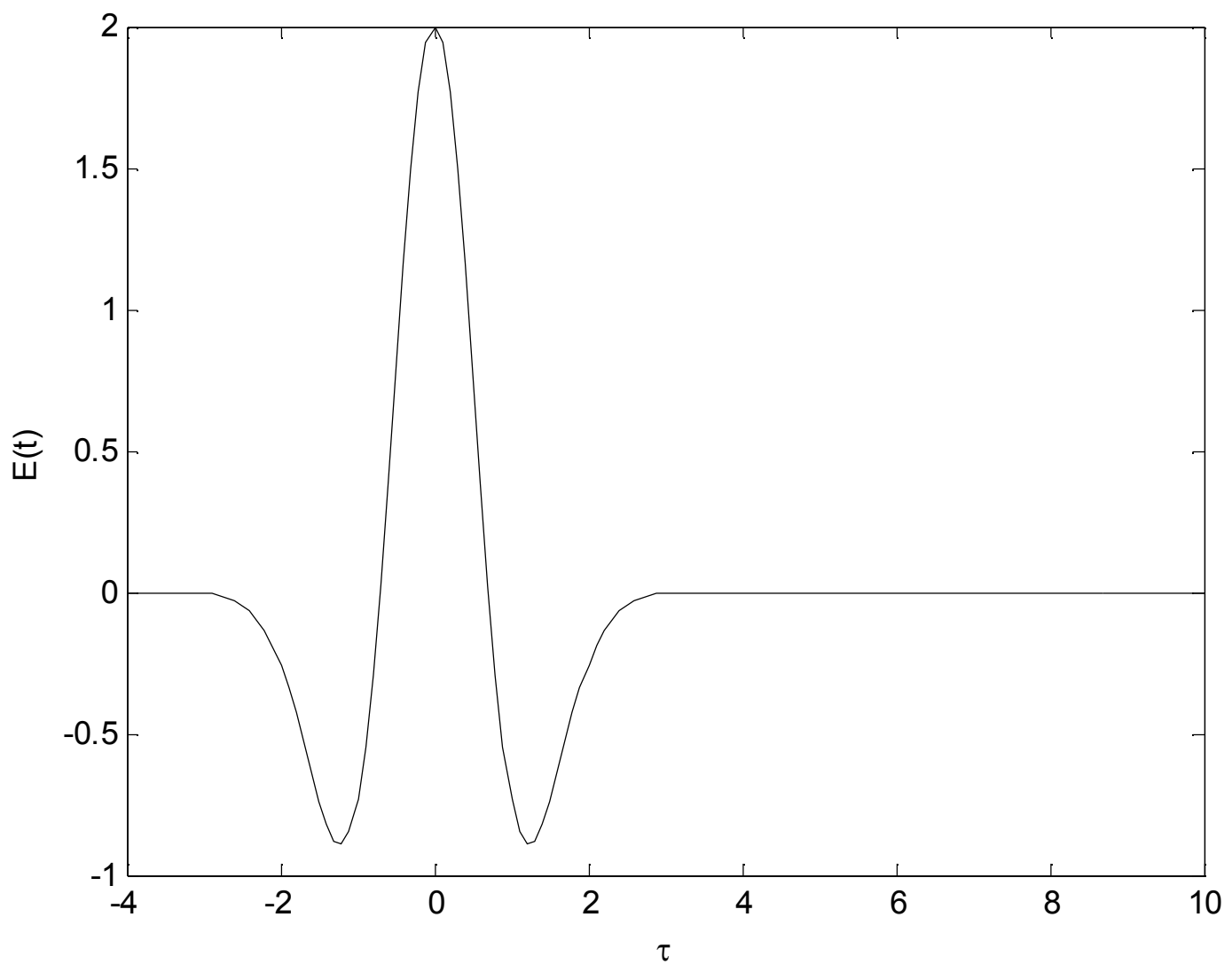

**Fig. 4.** Wavelets as derivatives of Gaussian function.

When describing wavelets of pulses [5] in the form of derivatives of the Gaussian function $u_0(0,t) = C_n \frac{d^n}{dt^n} e^{-t^2/\beta^2}$, then the field

$$u_P(x,t) = C_n \frac{d^n}{dt^n} \int_{-\infty}^{\infty} \exp\left[ -\left( t + \frac{x}{c}\theta + \frac{\rho}{c6}\theta^3 \right)^2 / \beta^2 \right] d\theta. \quad (12)$$

The results of calculations of the pulse shape near a caustic shown in Fig. 5.

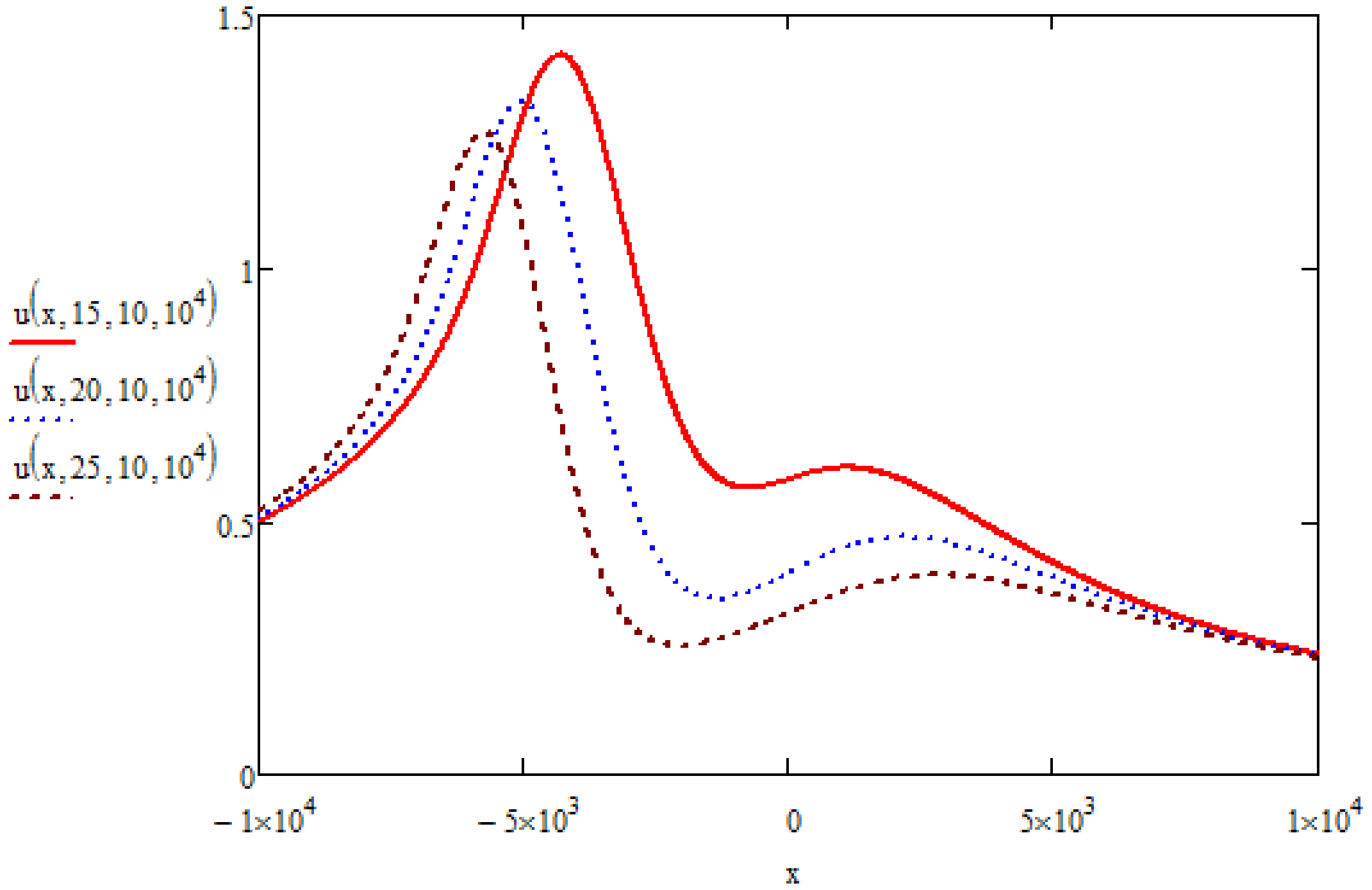

**Fig. 5.** The spatial distribution of field near the caustic for different moments of time: solid line – t=15 at.u., dotted line – t=20 at.u., dashed line – t = 25 at.u., pulse duration 10 at.u., abscissa is in at.u. .

## Conclusions

The diffraction of ultrashort pulses differs from the case of a monochromatic wave in dynamic change of a shadow. Spectral representation displays the shadows of different depths for different spectral components. In the scalar diffraction approximation to a fully absorbing screen problem is solved on the basis of the Huygens-Fresnel principle. To describe the temporal dynamics of the pulse diffracted at the edge of the half-plane, the most appropriate is a direct application of unsteady Kirchhoff-Sommerfeld integral. In this case it is possible to obtain a simple expression for the generalized solution in the form of a response to a single delta pulse. Diffraction the pulse of arbitrary shape is described by convolution the response function with a time-dependent initial pulse.

A feature of the pulse propagation in the caustic is the independence of the changes of the field components corresponding to different values of local angle parameter of turning radius of the caustic. It allows to get a unified integral representation for field the pulse of arbitrary shape.

The research was supported by the Russian Foundation for Basic Research (grant No. 015-07-09123 A) and by the Government order of the RF Ministry of Education and Science (project No. 1940).

## Rereferences